\documentclass[preprint,showpacs,preprintnumbers,amsmath,amssymb
              ,prb,eqsecnum]{revtex4}

\usepackage{graphicx}
\usepackage{dcolumn}
\usepackage{bm}

\begin{document}

\preprint{Sinv.ver2}

\title{$1/S$-expansion study of spin waves in a two-dimensional
Heisenberg antiferromagnet}

\author{Jun-ichi Igarashi}%
\affiliation{%
Faculty of Science, Ibaraki University, Mito, Ibaraki 310-8512, Japan}

\author{Tatsuya Nagao}
\affiliation{%
Faculty of Engineering, Gunma University, Kiryu, Gunma 376-8515, Japan}

\date{\today}

\begin{abstract}
We study the effects of quantum fluctuations on excitation spectra
in the two-dimensional Heisenberg antiferromagnet by means of the $1/S$
expansion. We calculate the spin-wave dispersion
and the transverse dynamical structure factor up to the second order of $1/S$
in comparison with inelastic neutron scattering experiments.
The spin-wave energy at momentum $(\pi,0)$ is found to be about $2\%$ 
smaller than that at $(\pi/2,\pi/2)$ due to the second-order correction.
In addition, we study the dimensional crossover from two dimensions 
to one dimension by weakening exchange couplings in one direction.
It is found that the second-order correction becomes large with approaching 
the quasi-one dimensional situation and makes the spin-wave energy 
approach to the des Cloizeaux-Pearson boundary for $S=1/2$. 
The transverse dynamical structure factor is also calculated up to the
second order of $1/S$. It is shown that the intensity of spin-wave peak
is strongly reduced while the intensity of three-spin-wave continuum 
becomes large and exceeds that of the spin-wave peak
in the quasi-one dimensional situation.
\end{abstract}

\maketitle

\section{\label{sec:1}Introduction}
The physics of low-dimensional spin systems has attracted much 
interest for past decades.
Although the quantum fluctuation is expected to be large in two dimensions,
there exist strong evidences that the quantum Heisenberg antiferromagnet
(QHAF) with nearest-neighbor coupling in a square lattice
exhibits the N\'eel long-range order at zero temperature.\cite{Chakravarty89}
Under the presence of the long-range order, the linear spin-wave (LSW) 
theory works rather well.\cite{Anderson52,Kubo52}
A natural way to include quantum fluctuations is
an expansion in terms of $1/S$, where $S$ is the magnitude of spin,
because the LSW theory is made up of a leading-order in the $1/S$ expansion.
Such attempts have been done and turned out to be 
useful.\cite{Oguchi60,Igarashi92-1,Igarashi92-2,Igarashi93,Canali92,Hamer92}

In our previous paper,\cite{Igarashi92-1} basing on the Holstein-Primakoff 
transformation,\cite{Holstein40} 
we calculated corrections up to the second order 
of $1/S$ in various physical quantities such as the spin-wave dispersion,
the sublattice magnetization, the perpendicular susceptibility,
and the spin-stiffness constant.
We also calculated the dynamical structure
factors of the transverse and the longitudinal components up to the 
second order of $1/S$.\cite{Igarashi92-2}
At that time, however, available experimental data of inelastic 
neutron scattering (INS) were limited to the momentum transfer 
in a narrow region of the Brillouin zone (BZ).
Therefore, our study was just a demonstration of usefulness
of the $1/S$ expansion.
Now that INS experiments provide us the information of the excitation 
spectra in the whole BZ, it may be interesting to calculate the excitation
spectra in the whole BZ in comparison with recent experiments.
We show that the second-order correction makes the spin wave energy
at momentum $(\pi,0)$ about $2\%$ smaller than that at $(\pi/2,\pi/2)$.
This difference is smaller than the value $7\sim 9\%$ obtained by
the series expansion\cite{Singh95,Zheng04} 
and the Monte Carlo simulation.\cite{Sandvik01}
We have so far not been able to find why the $1/S$ expansion 
within the second order gives different results, since the higher-order 
corrections is expected to be quite small. 
Note that the spin-wave dispersion was recently measured by
the INS experiment for Cu(DCOO)$_2\cdot$4D$_2$O (CFTD),
revealing the $6\%$ difference.\cite{Ronnow01,Christensen04}
This material is believed to be well described by the $S=1/2$
Heisenberg model within the nearest-neighbor coupling.\cite{Com1}
In addition to the spin-wave energy, we calculate the transverse dynamical 
structure factor up to the second order of $1/S$. 
It consists of the $\delta$-function-like peak
of one spin-wave excitation and the continuum of three spin-wave excitations.
The second-order correction is found quite small in the spin-wave-peak
intensity due to a cancellation of various second-order processes.
The result is compared with the recent experiment\cite{Christensen04} 
as well as other $1/S$-expansion study\cite{Canali93} based on the 
Dyson-Maleev transformation and the series expansion 
studies.\cite{Singh95,Zheng04}
In the three-spin-wave continuum, such a cancellation in the second-order
processes is not severe, and the substantial intensities come out.
This is consistent with our previous study\cite{Igarashi92-2}
and others.\cite{Canali93,Singh95,Zheng04}

Another purpose of this paper is to study the crossover behavior in
the QHAF from two dimensions to one dimension with weakening 
exchange coupling in one direction.  In purely one dimension,
of course, the antiferromagnetic long-range order
disappears due to quantum fluctuation 
and therefore the concept of spin waves breaks down.
Carrying out the $1/S$ expansion, we demonstrate that
the second-order corrections increase with approaching
the quasi-one dimensional situation.
The second-order correction works to increase considerably 
the sublattice magnetization, although the first-order correction makes
it decrease, in the quasi-one dimensional situation. 
Interestingly, the spin-wave dispersion is found to approach to the curve
known as the des Cloizaux-Pearson boundary in the $S=1/2$ QHAF 
chain.\cite{desCloizeaux62}
At the same time, for the spin-wave peak in the transverse dynamical 
structure factor, the first-order correction makes the intensity
decrease but the second-order correction makes it increase.
In the quasi-one dimensional situation, the former is much larger than
the latter, and the net intensity is strongly reduced from the LSW value.
On the other hand, the intensity of three-spin-wave continuum 
by the second-order correction increases and exceeds the spin-wave-peak
intensities in the quasi-one dimensional situation. 
This contrasts with the description of using spinon\cite{Ho01}
to describe the large intensity of the spectral continuum.
The above three characteristics in the quasi-one dimension, 
(a) the spin-wave energy approaches to the des Cloizaux-Pearson boundary,
(b) the spin-wave-peak intensity decreases, and (c) the intensity of
three-spin-wave continuum increases, are suggesting a close relation to
the purely-one dimensional behavior that the spectra are described by
continuum of two spinons.

The present paper is organized as follows.
In Sec. \ref{sect.2}, the Heisenberg Hamiltonian is expressed in terms of 
the 1/S-expansion. The Green's functions for spin waves are introduced
in Sec. \ref{sect.3}. The sublattice magnetization is calculated 
with the help of the Green's functions in Sec. \ref{sect.4}.
The spin-wave dispersion is calculated in Sec. \ref{sect.5},
and the transverse dynamical structure factor is calculated
in Sec. \ref{sect.6}.
Section \ref{sect.7} is devoted to the concluding remarks.

\section{\label{sect.2}Hamiltonian}

We consider the Heisenberg Hamiltonian on the square lattice
with directional anisotropy of exchange couplings:
\begin{equation}
 H= J\sum_{\bf\ell}{\bf S}_{\bf\ell}\cdot{\bf S}_{{\bf\ell}+{\bf a}}
  + J'\sum_{\bf\ell}{\bf S}_{\bf\ell}\cdot{\bf S}_{{\bf\ell}+{\bf b}},
\label{eq.heis}
\end{equation}
where ${\bf\ell}$ runs over all lattice sites and
${\bf\ell}+{\bf a}$ and ${\bf\ell}+{\bf b}$ indicate
the nearest neighbors to the ${\bf\ell}$th site in the positive $x$ and $y$
directions, respectively. 
Quasi-one dimensional situations are realized by weakening the exchange 
coupling $J'$ in the $y$ direction.

Introducing the Holstein-Primakoff transformation,\cite{Holstein40}
we express the spin operators in terms of boson annihilation operators 
$a_i$ and $b_j$ (and their Hermite conjugates),
\begin{eqnarray}
 S_i^z &=& S - a_i^\dagger a_i ,  
 \label{eq.boson1}\\
 S_i^+ &=& (S_i^-)^\dagger = \sqrt{2S}f_i(S)a_i ,\\
 S_j^z &=& -S + b_j^\dagger b_j ,\\
 S_j^+ &=& (S_j^-)^\dagger = \sqrt{2S}b_j^\dagger f_j(S) ,
 \label{eq.boson2}
\end{eqnarray}
where
\begin{equation}
    f_\ell (S) = \left(1 - \frac{n_\ell}{2S}\right)^{1/2}\\
               = 1 - \frac{1}{2}\frac{n_\ell}{2S} -
                  \frac{1}{8}\left(\frac{n_\ell}{2S}\right)^2 + \cdots .
\end{equation}
with $n_\ell=a_i^\dagger a_i$ and $b_j^\dagger b_j$.
Indices $i$ and $j$ refer to sites on the "up" and "down" sublattices, 
respectively.
Substituting Eqs.~(\ref{eq.boson1})-(\ref{eq.boson2}) into Eq.~(\ref{eq.heis})
we expand the Hamiltonian in powers of $1/S$ as
\begin{equation}
 H = -S^2N(J+J')+H_0 + H_1 + H_2 + \cdots, 
\end{equation}
with $N$ the number of lattice sites.

The leading term $H_0$ is expressed as
\begin{eqnarray}
 H_0 &=& JS\sum_{i}(2 a_i^\dagger a_i 
           + 2 b_{i+{\bf a}} b_{i+{\bf a}}
           +  a_i b_{i+{\bf a}}
           +  a_i b_{i-{\bf a}}
           + a_i^\dagger b_{i+{\bf a}}^\dagger
           + a_i^\dagger b_{i-{\bf a}}^\dagger ) \nonumber \\
     &+& J'S\sum_{i} (2 a_i^\dagger a_i 
           + 2 b_{i+{\bf b}} b_{i+{\bf b}}
           +  a_i b_{i+{\bf b}}
           +  a_i b_{i-{\bf b}}
           + a_i^\dagger b_{i+{\bf b}}^\dagger
           + a_i^\dagger b_{i-{\bf b}}^\dagger ).
\label{eq.h0}
\end{eqnarray}   
We diagonalize $H_0$ by rewriting the boson operators in the momentum 
space as
\begin{eqnarray}
 a_i &=& \left(\frac{2}{N}\right)^{1/2}\sum_{\bf k} a_{\bf k}
  \exp(i{\bf k\cdot r}_i), \\
 b_j &=& \left(\frac{2}{N}\right)^{1/2}
       \sum_{\bf k} b_{\bf k} \exp(i{\bf k\cdot r}_j),
\end{eqnarray}
and by introducing the Bogoliubov transformation,
\begin{equation}
 a_{\bf k}^\dagger = \ell_{\bf k}\alpha_{\bf k}^\dagger
          + m_{\bf k}\beta_{-{\bf k}}, \quad
 b_{-{\bf k}} = m_{\bf k}\alpha_{\bf k}^\dagger
          + \ell_{\bf k}\beta_{-{\bf k}}, 
\end{equation}
where
\begin{equation}
 \ell_{\bf k} = \Bigl[\frac{1+\epsilon_{\bf k}}
 {2\epsilon_{\bf k}}\Bigr]^{1/2},\quad
  m_{\bf k} = -{\rm sgn}(\gamma_{\bf k})\Bigl[\frac{1-\epsilon_{\bf k}}
  {2\epsilon_{\bf k}}\Bigr]^{1/2} \equiv - x_{\bf k}\ell_{\bf k},
\end{equation}
with
\begin{eqnarray}
 \epsilon_{\bf k} &=& \left(1-\gamma_{\bf k}^2\right)^{1/2},\\
   \gamma_{\bf k} &=& (\cos k_x+\zeta\cos k_y)/(1+\zeta), \\
           \zeta  &=& J'/J. 
\end{eqnarray}
Momentum ${\textbf k}$ is defined in the first magnetic Brillouin zone
(BZ). The ${\rm sgn}(\gamma_{\textbf k})$ denotes the sign of 
$\gamma_{\textbf k}$,
which is absorbed into the definition of $x_{\bf k}$.
For the study of the isotropic exchange coupling ($\zeta=1$), 
we have neglected this factor 
because $\gamma_{\bf k}$ is always positive in the first BZ.
For the anisotropic coupling, however, this redefinition of $x_{\bf k}$ 
is necessary, because $\gamma_{\bf k}$ is negative in a certain region 
of the first BZ. After this transformation, we have
\begin{eqnarray}
 H_0 &=& 2JS(1+\zeta)\sum_{\bf k}(\epsilon_{\bf k}-1) \nonumber \\
     &+& 2JS(1+\zeta)\sum_{\bf k} \epsilon_{\bf k}
   (\alpha_{\bf k}^\dagger \alpha_{\bf k}
   + \beta_{\bf k}^\dagger\beta_{\bf k}). 
\end{eqnarray}
This expression is the same as that for the isotropic coupling,
except for the first factor $2JS(1+\zeta)$.

The first-order term $H_1$ can be expressed in terms of 
spin-wave operators through the same procedure as above.
The result for the anisotropic coupling is given by
the previous expression in Ref.~\onlinecite{Igarashi92-1} with 
simply replacing 
$JSz$ by $2JS(1+\eta)$:
\begin{eqnarray}
 H_1 &=& \frac{2JS(1+\zeta)}{2S} A\sum_{\bf k}\epsilon_{\bf k}
 (\alpha_{\bf k}^\dagger \alpha_{\bf k} + \beta_{\bf k}^\dagger\beta_{\bf k})
 \nonumber \\
     &+&\frac{-2JS(1+\zeta)}{2SN}\sum_{1234}\delta_{\bf G}(1+2-3-4)
     \ell_1\ell_2\ell_3\ell_4  \nonumber \\
     &\times& \biggl[\alpha_1^\dagger\alpha_2^\dagger\alpha_3\alpha_4 
     B_{1234}^{(1)}+\beta_{-3}^\dagger\beta_{-4}^\dagger\beta_{-1}\beta_{-2} 
     B_{1234}^{(2)}+4\alpha_1^\dagger\beta_{-4}^\dagger\beta_{-2}\alpha_3 
     B_{1234}^{(3)}  \nonumber \\
     &+&\bigl(2\alpha_1^\dagger\beta_{-2}\alpha_3\alpha_4 B_{1234}^{(4)}
      +2\beta_{-4}^\dagger\beta_{-1}\beta_{-2}\alpha_3 B_{1234}^{(5)}
      +\alpha_1^\dagger\alpha_2^\dagger\beta_{-3}^\dagger\beta_{-4}^\dagger
     B_{1234}^{(6)} + {\rm H.c.}\bigr)\biggr],
\label{eq.intham}
\end{eqnarray}
with 
\begin{equation}
 A=\frac{2}{N}\sum_{\bf k}(1-\epsilon_{\bf k}) .
 \label{eq.A}
\end{equation}
Momenta ${\bf k}_1$, ${\bf k}_2$, ${\bf k}_3$, $\cdots$ are abbreviated
as $1,2,3,\cdots$.
The first term arises from setting the products of four boson operators 
into normal product forms with respect to spin-wave operators.
The second term in Eq.~(\ref{eq.intham}) represents the scattering
of spin waves. The Kronecker delta $\delta_{\bf G}(1+2-3-4)$ represents
the conservation of momenta within a reciprocal lattice vector ${\bf G}$.
The vertex functions $B^{(i)}$'s in a symmetric
parameterization are the same as those given by Eqs.~(2.16)-(2.20) 
in Ref.~\onlinecite{Igarashi92-1}, so that they are omitted here.

The second-order term $H_2$ is composed of products of six boson operators.
Writing it in a normal product form with respect to spin-wave operators,
we have 
\begin{equation}
 H_2 = \frac{2JS(1+\zeta)}{(2S)^2}\sum_{\bf k}\left[ C_1({\bf k})
  (\alpha_{\bf k}^\dagger \alpha_{\bf k} + \beta_{\bf k}^\dagger\beta_{\bf k})
       + C_2({\bf k})(\alpha_{\bf k}^\dagger\beta_{\bf -k}^\dagger
             + \beta_{\bf -k}\alpha_{\bf k})+\cdots \right].
\end{equation}
Neglected terms are unnecessary for calculating corrections up to the
second order. The explicit forms of $C_1({\bf k})$ and $C_2({\bf k})$,
are given by Eqs.~(2.22) and (2.23) in Ref.~\onlinecite{Igarashi92-1}.
Note that $C_1({\bf k})$ and $C_2({\bf k})$ diverge as $1/\epsilon_{\bf k}$ 
with $|{\bf k}|\to 0$. 
       
\section{\label{sect.3}Green's Function} 

For systematically carrying out the 1/S-expansion,
it is convenient to introduce the Green's functions for spin-waves,
\begin{eqnarray}
 G_{\alpha\alpha}({\bf k},t) &=& -i \langle T(\alpha_{\bf k}(t)
 \alpha_{\bf k}^\dagger (0)) \rangle, \\
 G_{\alpha\beta}({\bf k},t) &=& -i \langle T(\alpha_{\bf k}(t)
 \beta_{-{\bf k}}(0)) \rangle, \\
 G_{\beta\alpha}({\bf k},t) &=& -i \langle T(\beta_{-{\bf k}}^\dagger(t)
 \alpha_{\bf k}^\dagger (0)) \rangle, \\
 G_{\beta\beta}({\bf k},t) &=& -i \langle T(\beta_{-{\bf k}}^\dagger(t)
 \beta_{-{\bf k}}(0)) \rangle,
\end{eqnarray}
where $ \langle \cdots \rangle$ denotes the average over the ground state, and 
{\textit T} is the time-ordering operator.

In this paper, we measure energies in units of $2JS(1+\zeta)$.
The unperturbed propagators corresponding to $H_0$ are given by
\begin{eqnarray}
 G_{\alpha\alpha}^0({\bf k},\omega) &=& [\omega 
 - \epsilon_{\bf k} + i\delta]^{-1}, \\
 G_{\alpha\beta}^0({\bf k},\omega) &=& G_{\beta\alpha}^0({\bf k},\omega) = 0, 
 \\
 G_{\beta\beta}^0({\bf k},\omega) &=& [-\omega 
 - \epsilon_{\bf k} + i\delta]^{-1}.
\end{eqnarray}
The self-energy is defined by
\begin{equation}
 G_{\mu\nu}({\bf k},\omega) = G_{\mu\nu}^0({\bf k},\omega)
 + \sum_{\mu'\nu'} G_{\mu\mu'}^0({\bf k},\omega)
 \Sigma_{\mu'\nu'}({\bf k},\omega)G_{\nu'\nu}({\bf k},\omega). 
\end{equation}
It is expanded in powers of $1/(2S)$,
\begin{equation}
 \Sigma_{\mu\nu}({\bf k},\omega) = \frac{1}{2S}
    \Sigma_{\mu\nu}^{(1)}({\bf k},\omega)
 + \frac{1}{(2S)^2}\Sigma_{\mu\nu}^{(2)}({\bf k},\omega) + \cdots. 
\end{equation}
>From $H_1$ we have the first-order correction as
\begin{equation}
 \Sigma_{\alpha\alpha}^{(1)}({\bf k},\omega)
  = \Sigma_{\beta\beta}^{(1)}({\bf k},\omega) = A\epsilon_{\bf k}, \quad
 \Sigma_{\alpha\beta}^{(1)}({\bf k},\omega)
   = \Sigma_{\beta\alpha}^{(1)}({\bf k},\omega)= 0.
\end{equation}

\begin{figure}
\includegraphics[width=8.0cm]{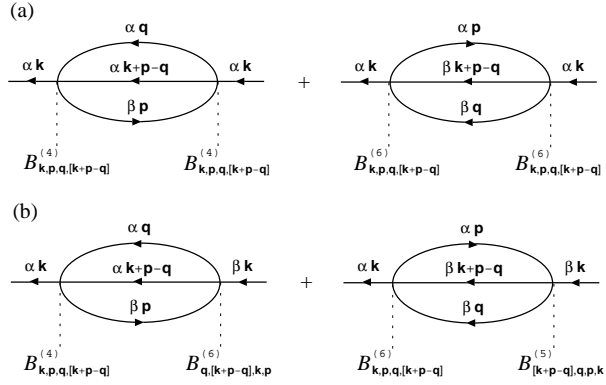}%
\caption{\label{fig.diagram1}
Second-order diagrams for the self-energy, 
(a) $\Sigma_{\alpha\alpha}({\bf k},\omega)$ 
and (b) $\Sigma_{\alpha\beta}({\bf k},\omega)$.
Solid lines represent the unperturbed Green's functions.
}
\end{figure}

The second-order term $\Sigma_{\mu\nu}^{(2)}({\bf k},\omega)$ is obtained from
the second-order perturbation, whose diagrams are shown 
in Fig.~\ref{fig.diagram1}.
We obtain formally the same expression for the self-energy as 
in our previous paper:\cite{Igarashi92-1}
\begin{eqnarray}
 \Sigma_{\alpha\alpha}^{(2)}({\bf k},\omega)
  &=& \Sigma_{\beta\beta}^{(2)}(-{\bf k},-\omega) \nonumber \\
  &=& C_1({\bf k}) + \left(\frac{2}{N}\right)^2
  \sum_{{\bf p}{\bf q}}2\ell_{\bf k}^2\ell_{\bf p}^2
  \ell_{\bf q}^2\ell_{{\bf k}+{\bf p}-{\bf q}}^2  \nonumber \\
 &\times& \left[\frac{\mid B_{{\bf k},{\bf p},{\bf q},[{\bf k+p-q}]}^{(4)}
 \mid^2}{\omega-\epsilon_{\bf p}-\epsilon_{\bf q}-\epsilon_{\bf k+p-q}+i\delta}
 -\frac{\mid B_{{\bf k},{\bf p},{\bf q},[{\bf k+p-q}]}^{(6)}\mid^2}
{\omega+\epsilon_{\bf p}+\epsilon_{\bf q}+\epsilon_{\bf k+p-q}-i\delta}\right], 
\label{eq.self1}\\
 \Sigma_{\alpha\beta}^{(2)}({\bf k},\omega)
  &=& \Sigma_{\beta\alpha}^{(2)}(-{\bf k},-\omega) \nonumber \\
  &=& C_2({\bf k}) + \left(\frac{2}{N}\right)^2
  \sum_{{\bf p}{\bf q}}2
  \ell_{\bf k}^2\ell_{\bf p}^2\ell_{\bf q}^2\ell_{{\bf k}+
 {\bf p}-{\bf q}}^2 {\rm sgn}(\gamma_{\bf G}) \nonumber \\
 &\times& B_{{\bf k},{\bf p},{\bf q},[{\bf k+p-q}]}^{(4)}
          B_{{\bf k},{\bf p},{\bf q},[{\bf k+p-q}]}^{(6)}
 \frac{2(\epsilon_{\bf p}+\epsilon_{\bf q}+\epsilon_{{\bf k}+{\bf p}-{\bf q}})}
 {\omega^2-(\epsilon_{\bf p}+\epsilon_{\bf q}+\epsilon_{\bf k+p-q})^2+i\delta},
\label{eq.self2}
\end{eqnarray}
where $\delta\to 0$, and $[{\bf k+p-q}]$ is the vector ${\bf k+p-q}$ reduced to
the 1st BZ by a reciprocal vector ${\bf G}$.
We have used the relations
\begin{equation}
\begin{array}{l}
B_{[{\bf k+p-q}],{\bf q},{\bf p},{\bf k}}^{(5)} =
{\rm sgn}(\gamma_{\bf G})B_{{\bf k},{\bf p},{\bf q},[{\bf k+p-q}]}^{(4)},\\
B_{{\bf q},[{\bf k+p-q}],{\bf k},{\bf p}}^{(6)} =
{\rm sgn}(\gamma_{\bf G})B_{{\bf k},{\bf p},{\bf q},[{\bf k+p-q}]}^{(6)}.  \\
\end{array}
\end{equation}
The terms divergent with ${\bf k}\to 0$ in $C_1({\bf k})$ and $C_2{\bf k})$
are canceled by the second-order perturbation terms in Eqs.~(\ref{eq.self1}) 
and (\ref{eq.self2}).
One can prove $\Sigma_{\mu\nu}^{(2)}({\bf k}\to 0,\omega=0)\to 0$
from these equations.

\section{\label{sect.4}Sublattice Magnetization} 

Once the Green's function is known, the sublattice magnetization is 
calculated from the relation,
\begin{eqnarray}
 M &\equiv& S-\langle a_i^\dagger a_i\rangle \nonumber \\
   &=& S - \frac{2}{N}\sum_{\bf k}\lim_{\eta\to 0^+}\int_{-\infty}^{+\infty}
\frac{{\rm d}\omega}{2\pi}i{\rm e}^{i\omega\eta}
\Bigl\{\ell_{\bf k}^2G_{\alpha\alpha}({\bf k},\omega) \nonumber\\
&+& \ell_{\bf k}m_{\bf k}\bigl[G_{\alpha\beta}({\bf k},\omega)
+G_{\beta\alpha}({\bf k},\omega)\bigr]
+ m_{\bf k}^2G_{\beta\beta}({\bf k},\omega)\Bigr\},
\end{eqnarray}
with $\eta\to 0^+$.
After carrying out the integration with respect to $\omega$,
we obtains
\begin{equation}
 M = S -\Delta S + \frac{M_2}{(2S)^2},
\end{equation}
with
\begin{eqnarray}
 \Delta S &=& \frac{2}{N}\sum_{\bf k}\frac{1}{2}(\epsilon_{\bf k}^{-1}-1), \\
 M_2 &=& \frac{2}{N}\sum_{\bf k}\Biggl\{
 \frac{\ell_{\bf k}m_{\bf k}}{\epsilon_{\bf k}}\Sigma_{\alpha\beta}^{(2)}
({\bf k},-\epsilon_{\bf k}) \nonumber\\
 &-& \left(\frac{2}{N}\right)^2\sum_{{\bf pq}}
 2\ell_{\bf k}^2\ell_{\bf p}^2\ell_{\bf q}^2\ell_{\bf k+p-q}^2
 \biggl[\frac{(\ell_{\bf k}^2+m_{\bf k}^2)\mid
 B_{{\bf k},{\bf p},{\bf q},[{\bf k+p-q}]}^{(6)}\mid^2}
 {(\epsilon_{\bf k}+\epsilon_{\bf p}+\epsilon_{\bf q}+\epsilon_{\bf k+p-q})^2} 
 \nonumber\\
 &+&\frac{2\ell_{\bf k} m_{\bf k}{\rm sgn}(\gamma_{\bf G})
 B_{{\bf k},{\bf p},{\bf q},[{\bf k+p-q}]}^{(4)}
 B_{{\bf k},{\bf p},{\bf q},[{\bf k+p-q}]}^{(6)}}
 {\epsilon_{\bf k}^2-(\epsilon_{\bf p}+\epsilon_{\bf q}
 +\epsilon_{\bf k+p-q})^2}\biggr]\Biggr\}.
\label{eq.zero}
\end{eqnarray}
Here $[{\bf k+p-q}]$ was defined before, and 
${\bf G}={\bf k+p-q}-[{\bf k+p-q}]$. The zeroth-order correction
$\Delta S$ represents the well-known ``zero-point" reduction in the LSW
theory. 
To evaluate Eq.~(\ref{eq.zero}), we sum up the values of $N_L^2/4$ points
of ${\bf k}$ in the 1/4 part of the first BZ and $N_L^2$ points of
${\bf p}$ and ${\bf q}$ in the first BZ, with $N_L=20,48$.
For $J'/J=1$, the convergence is very good; $M_2=0.0035059$ for $N_L=20$,
and $M_2=0.0035065$ for $N_L=48$.

\begin{table}
\caption{\label{table2}
Sublattice magnetization}
\begin{ruledtabular}
\begin{tabular}{llllll}
 $J'/J$    & $1$ & $0.5$ & $0.1$ & $0.075$ & $0.05$ \\
\hline
 $\Delta S$ & $0.196$  & $0.213$ & $0.355$ & $0.391$ & $0.445$ \\ 
 $M_2$      & $0.0035$ & $0.024$ & $0.323$ & $0.481$ & $0.818$ \\ 
\end{tabular}
\end{ruledtabular}
\end{table}

Table \ref{table2} lists the values of $\Delta S$ and $M_2$ for several 
values of $J'/J$. These values are evaluated for $N_L=48$.
For the isotropic coupling, we reproduce the values obtained previously.
\cite{Igarashi92-1} 
The zero-point reduction $\Delta S$ increases with decreasing values of $J'/J$.
On the other hand, $M_2$ is found always positive, tending to cancel
the zero-point reduction. The value increases with decreasing values of
$J'/J$, and becomes comparable to $\Delta S$ around 
$J'/J=0.1$, suggesting the applicability limit of the expansion for 
the case of $S=1/2$. 

\section{\label{sect.5}Spin-Wave Dispersion} 

Within the second order in $1/S$,
the renormalized spin-wave energy
${\tilde \epsilon_{\bf k}}$ in units of $2JS(1+\zeta)$
is obtained from 
\begin{equation}
 {\tilde\epsilon}_{\bf k} = \epsilon_{\bf k} + \frac{1}{2S}A\epsilon_{\bf k} 
 + \frac{1}{(2S)^2}\Sigma_{\alpha\alpha}^{(2)}({\bf k},\epsilon_{\bf k}).
\label{eq.spin-wave}
\end{equation}
From this equation, we define the renormalized spin-wave velocity $V_x$ 
along the $x$ direction by $V_x \equiv \lim_{k_x\to 0}
2JS(1+\zeta)\tilde \epsilon_{\bf k}/(\hbar k_x)$ with $k_y=0$.
Thus the renormalization factor is expressed as
\begin{equation}
 Z_v\equiv\frac{V_x}{2JS(1+\zeta)^{1/2}}= 1+\frac{v_1}{2S}+\frac{v_2}{(2S)^2}, 
\end{equation}
with $v_1=A$ given by Eq.~(\ref{eq.A}).
In the following numerical evaluation, we divide
the first BZ into $N_L^2$ meshes with $N_L=64$.

\subsection{Isotropic case}

\begin{figure}
\includegraphics[width=8.0cm]{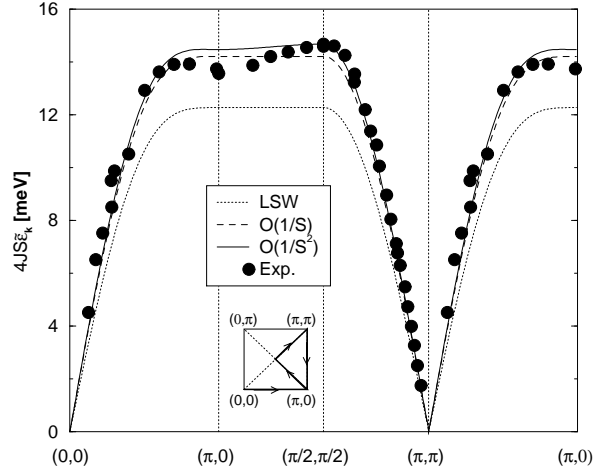}%
\caption{\label{fig.iso.disp}
Spin-wave energy as a function of momentum for the isotropic coupling
($J'/J=1$). $S=1/2$.
The dotted, broken and solid lines represent the values calculated 
within the LSW theory, up to the first-order correction, 
and up to the second-order correction, respectively. 
Experimental data are taken from Ref.~\onlinecite{Christensen04}.
Inset indicates high symmetry lines which momentum varies along.
}
\end{figure}

Figure \ref{fig.iso.disp} shows the spin-wave energy 
$2JS(1+\zeta)\tilde\epsilon_{\bf k}$
as a function of momentum for the isotropic coupling ($J'/J=1$)
with $S=1/2$, in comparison with the experimental data
taken from the INS for 
Cu(DCOO)$_2\cdot$4D$_2$O (CFTD).\cite{Ronnow01,Christensen04}
Momentum is measured in units of (nearest neighbor distance)$^{-1}$.
In the whole BZ, both the first and second order corrections make
the spin-wave energy larger.
The curve along the line $(0,0)-(\pi,0)$ has already been reported
in our previous paper.\cite{Igarashi92-1}
The dispersion along $(\pi/2,\pi/2)-(\pi,0)$ is completely 
flat within the first-order correction.
The second-order correction makes the excitation energy 
at $(\pi,0)$ about $2\%$ smaller than the energy at $(\pi/2,\pi/2)$.
Explicitly they are 
$\tilde\epsilon_{(\pi/2,\pi/2)}=1.196$ and $\tilde\epsilon_{(\pi,0)}=1.179$. 
A previous series expansion study predicted the energy difference 
about $7\%$,\cite{Singh95} and a recent study gave about $9\%$ difference,
that is, $\tilde\epsilon_{(\pi/2,\pi/2)}=1.192$ and 
$\tilde\epsilon_{(\pi,0)}=1.09$.\cite{Zheng04}
A Monte Carlo simulation has given 
$\tilde\epsilon_{(\pi/2,\pi/2)}=1.195$ and 
$\tilde\epsilon_{(\pi,0)}=1.08$.\cite{Sandvik01}
These values at $(\pi/2,\pi/2)$ agree well with our value, while
the values at $(\pi,0)$ is rather different from our estimate.
The experimental data indicate that the excitation energy at $(\pi,0)$ is 
$6\%$ smaller than that at $(\pi/2,\pi/2)$.\cite{Ronnow01,Christensen04}

\subsection{Anisotropic case}

\begin{figure}
\includegraphics[width=8.0cm]{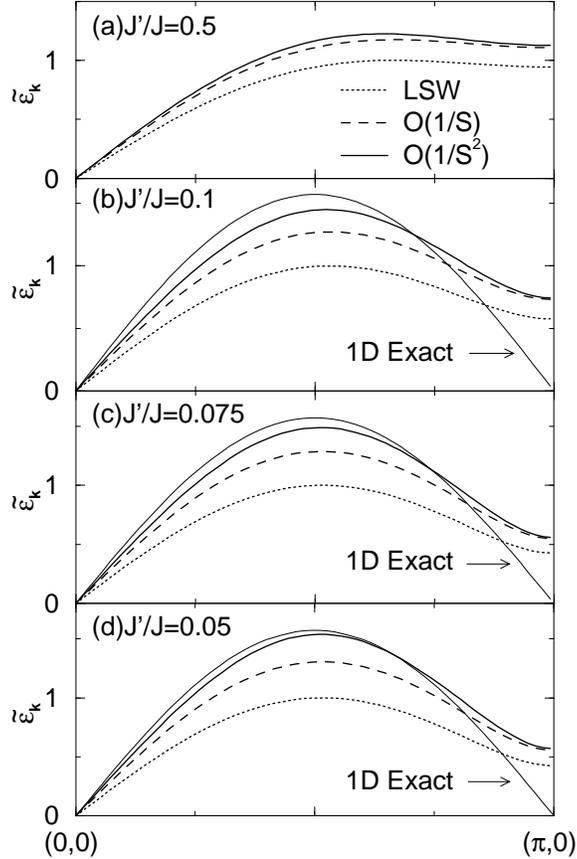}%
\caption{\label{fig.aniso.disp}
Spin-wave energy as a function of momentum for anisotropic couplings,
(a) $\zeta=0.5$, (b) $\zeta=0.1$, (c) $\zeta=0.075$, and (d) $\zeta=0.05$.
The dotted, broken and solid lines represent the values calculated 
within the LSW theory, up to the first-order correction, and up to
the second-order correction, respectively. 
The thin solid line labelled ``1D Exact" represents the des Cloizeaux-Pearson 
boundary.}
\end{figure}

Figure \ref{fig.aniso.disp} shows the renormalized spin-wave energy 
as a function of momentum along $(0,0)-(\pi,0)$ for $J'/J<1$.
As the same as the isotropic coupling,
both the first-order and the second-order 
corrections are found to be positive, making the energy larger. 
Both corrections increase with decreasing values of $J'/J$.
In quite small interchain couplings ($J'/J < 0.1$),
the excitation energy seems approaching 
the {\em des Cloizeaux-Pearson boundary} 
in one dimension.\cite{desCloizeaux62}
As shown in Table \ref{table1}, the renormalization constant
$Z_v$ seems approaching $\pi/2$, corresponding to the value of
the boundary.

\begin{table}
\caption{\label{table1}
Renormalization of spin-wave velocity}
\begin{ruledtabular}
\begin{tabular}{llllll}
 $J'/J$& $1$     & $0.5$   & $0.1$   & $0.075$ & $0.05$ \\
\hline
 $v_1$ & $0.158$ & $0.174$ & $0.272$ & $0.287$ & $0.306$ \\ 
 $v_2$ & $0.021$ & $0.053$ & $0.130$ & $0.141$ & $0.155$ \\ 
\end{tabular}
\end{ruledtabular}
\end{table}

\section{\label{sect.6}Dynamical Structure Factor} 

The dynamical structure factor is an important quantity, 
since it is directly related to the INS spectra.
We have already reported the expression within the second order 
of $1/S$ in the isotropic coupling situation.\cite{Igarashi92-2}
As is evident from the forms of $H_0$ and $H_1$, the
formulas for the anisotropic coupling are
formally the same as those for
the isotropic coupling.

We consider only the transverse component, 
which is defined by
\begin{equation}
 S^{+-}_{u(s)}({\bf k}, \omega) = \frac{1}{2\pi}
 \int {\rm d}t {\rm e}^{i\omega t}
   \langle Q_{u(s)}({\bf k},t)Q_{u(s)}({\bf k},0)^{\dagger}\rangle,
\end{equation}
where
\begin{equation}
 Q_{u(s)}({\bf k})=S^{+}_a({\bf k})\pm S^{+}_b({\bf k}),
 \label{eq.Q}
\end{equation}
with 
\begin{eqnarray}
 S^{+}_a({\bf k})&=&[S^{-}_a({\bf k})]^{\dagger}=
  \left(\frac{2}{N}\right)^{1/2}\sum_i S^{+}_i \exp(-i{\bf k}\cdot{\bf r}_i),
  \label{eq.sa}\\
 S^{+}_b({\bf k})&=&[S^{-}_b({\bf k})]^{\dagger}=
  \left(\frac{2}{N}\right)^{1/2}\sum_j S^{+}_j \exp(-i{\bf k}\cdot{\bf r}_j).
  \label{eq.sb}
\end{eqnarray}
We need the ``uniform" and ``staggered" parts because the momentum 
is defined inside the first BZ.
They are labelled as the suffix ``u" and ``s", 
and correspond to upper and lower signs in Eq.~(\ref{eq.Q}), respectively.

We start by introducing the operators,
\begin{eqnarray}
 Y^{+}_{\alpha}({\bf k})&=&[Y^{-}_{\alpha}({\bf k})]^{\dagger}
   = [\ell_{\bf k}S^{+}_{a}({\bf k})-m_{\bf k}S^{+}_{b}({\bf k})]/(2S)^{1/2},\\
 Y^{+}_{\beta}({\bf k})&=&[Y^{-}_{\beta}({\bf k})]^{\dagger}
   = [-m_{\bf k}S^{+}_{a}({\bf k})+\ell_{\bf k}S^{+}_{b}({\bf k})]/(2S)^{1/2},
\end{eqnarray}
and the associated Green's functions,
\begin{equation}
 F_{\mu\nu}({\bf k},\omega)=-i\int_{-\infty}^{\infty}dt {\rm e}^{i\omega t}
   \langle T[Y^{+}_{\mu}({\bf k},t)Y^{-}_{\nu}({\bf k},0)]\rangle.
\end{equation}
Then, with the help of the fluctuation-dissipation theorem,
we have
\begin{eqnarray}
& & S^{+-}_{u(s)}({\bf k},\omega) = 
2S(\ell_{\bf k}\pm m_{\bf k})^2 \left(- \frac{1}{\pi} \right) \nonumber\\
&  &\times {\rm Im}[F_{\alpha\alpha}({\bf k},\omega) 
                \pm F_{\alpha\beta}({\bf k},\omega)
                \pm F_{\beta\alpha}({\bf k},\omega)
                   +F_{\beta\beta}({\bf k},\omega)],
\nonumber \\
 \label{eq.dyna}
\end{eqnarray}
where the upper (lower) signs correspond to the uniform (staggered) part.

For calculating $F_{\mu\nu}({\bf k},\omega)$, we expand the operator 
$Y^{+}_{\mu}({\bf k})$ in terms of spin-wave operators with the help of
the HP transformation and the Bogoliubov transformation.
After lengthy calculations, we have
\begin{eqnarray}
 Y^{+}_{\alpha}({\bf k}) &=&
  D\alpha_{\bf k}-\frac{1}{2S}\frac{2}{N}\sum_{234}
    \delta_{\bf G}({\bf k}+2-3-4)\frac{1}{2}\ell_{\bf k}\ell_2\ell_3\ell_4
 \nonumber\\
 &\times& (M^{(1)}_{{\bf k}234}\beta_{-2}\alpha_3\alpha_4
          +M^{(2)}_{{\bf k}234}\alpha^{\dagger}_{2}\beta^{\dagger}_{-3}
           \beta^{\dagger}_{-4}+ \cdots),
  \label{eq.y1}\\
 Y^{+}_{\beta}({\bf k}) &=&
  D\beta^{\dagger}_{-\bf k} \nonumber \\
&-&\frac{1}{2S}\frac{2}{N}\sum_{234}
    \delta_{\bf G}({\bf k}+2-3-4)\frac{1}{2}\ell_{\bf k}\ell_2\ell_3\ell_4
    {\rm sgn}(\gamma_{\bf G})\nonumber\\
 &\times& (M^{(2)}_{{\bf k}234}\beta_{-2}\alpha_3\alpha_4
          +M^{(1)}_{{\bf k}234}\alpha^{\dagger}_{2}\beta^{\dagger}_{-3}
           \beta^{\dagger}_{-4}+ \cdots),
  \label{eq.y2}
\end{eqnarray}
where
\begin{eqnarray}
 D&=&1-\frac{\Delta S}{2S}-\frac{1}{4}\frac{\Delta S(1+3\Delta S)}{(2S)^2},
 \label{eq.D}\\
 M^{(1)}_{{\bf k}234} &=& -x_2+ {\rm sgn}(\gamma_{\bf G})x_{\bf k}x_3x_4, \\
 M^{(2)}_{{\bf k}234} &=& x_3x_4 - {\rm sgn}(\gamma_{\bf G})x_{\bf k}x_2.
\end{eqnarray}
with ${\bf G}={\bf k}+2-3-4$. 
The first-order and second-order corrections in Eq.~(\ref{eq.D}) arises from 
setting four and six boson operators in the HP transformation 
into the normal product forms with spin-wave operators, respectively.
Thereby, the second terms in Eqs. (\ref{eq.y1}) and (\ref{eq.y2})
are normally ordered.
Note that ${\rm sgn}(\gamma_{\bf G})$ arises from the phase difference
in the definitions, Eqs.~(\ref{eq.sa}) and (\ref{eq.sb}).

\begin{figure}
\includegraphics[width=8.0cm]{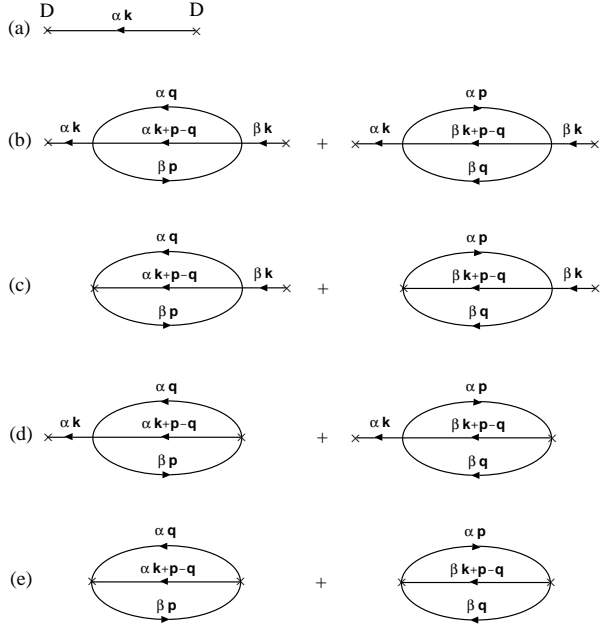}%
\caption{\label{fig.struc.diagram}
Diagrams for $F_{\mu\nu}({\bf k},\omega)$.
Solid lines represent the unperturbed Green's functions 
$G_{\mu\nu}^0({\bf k},\omega)$. 
}
\end{figure}

With the use of Eqs.~(\ref{eq.y1}) and (\ref{eq.y2}),
$F_{\mu\nu}({\bf k},\omega)$ is expanded up to the second order by
the diagrams shown in Fig.~\ref{fig.struc.diagram}.
Explicitly, it is given by
\begin{eqnarray}
 F_{\mu\nu}({\bf k},\omega) &=& D^2G_{\mu\nu}^0({\bf k},\omega)\delta_{\mu\nu}
   + G_{\mu\mu}^0({\bf k},\omega)\frac{1}{(2S)^2}
   \Sigma_{\mu\nu}^{(2)}({\bf k},\omega)G_{\nu\nu}^0({\bf k},\omega)\nonumber\\
  &+& I_{\mu\nu}({\bf k},\omega)G_{\nu\nu}^0({\bf k},\omega) 
   + G_{\mu\mu}^0({\bf k},\omega)\tilde I_{\mu\nu}({\bf k},\omega)
   + J_{\mu\nu}({\bf k},\omega),
 \label{eq.f}
\end{eqnarray}
where each term in Eq.~(\ref{eq.f}) corresponds to the diagrams 
(a), (b), (c), (d), and (e), respectively.
Explicit expressions for $I_{\mu\nu}({\bf k},\omega)$,
$\tilde I_{\mu\nu}({\bf k},\omega)$, and $J_{\mu\nu}({\bf k},\omega)$ 
are given by Eqs.~(4.5)-(4.22) in Ref.~\onlinecite{Igarashi92-1}.

The dynamical structure factor is obtained by substituting Eq.~(\ref{eq.f}) 
into Eq.~(\ref{eq.dyna}). 
It consists of the $\delta$-function-like peak of the one spin-wave excitation
and the continuum of three spin-wave excitations:
\begin{equation}
 S_{u(s)}^{+-}({\bf k},\omega) = 
   \rho_{u(s)}^{(1)}({\bf k})\delta(\omega-\epsilon_{\bf k})
                        + \rho_{u(s)}^{(2)}({\bf k},\omega).
 \label{eq.s+-}
\end{equation}
For the term of one spin-wave excitation, the bare energy $\epsilon_{\bf k}$
is to be replaced by the renormalized value $\tilde\epsilon_{\bf k}$
given by Eq.~(\ref{eq.spin-wave}).
However, the spectral weight $\rho^{(1)}({\bf k})$ 
within the second order of $1/S$ is safely evaluated 
by putting $\omega=\epsilon_{\bf k}$ in Eq.~(\ref{eq.f}).
It is expressed as
\begin{equation}
 \rho_{u(s)}^{(1)}({\bf k}) = 2S(\ell_{\bf k}\pm m_{\bf k})^2
 \left(1+\frac{d_{u(s),1}}{2S}+\frac{d_{u(s),2}}{(2S)^2}\right),
\end{equation}
with
\begin{eqnarray}
 d_{u(s),1} &=& -2\Delta S, 
 \label{eq.dyna1}\\
 d_{u(s),2} &=& -\frac{1}{2}\Delta S(1+\Delta S) 
             \mp\frac{1}{\epsilon_{\bf k}} 
             \Sigma_{\alpha\beta}^{(2)}({\bf k},\epsilon_{\bf k}) \nonumber\\
  &+& \left(\frac{2}{N}\right)^2\sum_{\bf pq}
     2\ell_{\bf k}^2\ell_{\bf p}^2\ell_{\bf q}^2\ell_{\bf k+p-q}^2\nonumber\\
  &\times&\left[\frac{-|B^{(4)}_{\bf k,p,q,[k+p-q]}|^2}
  {(\epsilon_{\bf k}-\epsilon_{\bf p}-\epsilon_{\bf q}-\epsilon_{\bf k+p-q})^2}
  +\frac{|B^{(6)}_{\bf k,p,q,[k+p-q]}|^2}
  {(\epsilon_{\bf k}+\epsilon_{\bf p}+\epsilon_{\bf q}+\epsilon_{\bf k+p-q})^2}
 \right] \nonumber\\
  &+& \left(\frac{2}{N}\right)^2\sum_{\bf pq}
     2\ell_{\bf k}^2\ell_{\bf p}^2\ell_{\bf q}^2\ell_{\bf k+p-q}^2
 (M_{\bf k,p,q,[k+p-q]}^{(1)}
     \pm{\rm sgn}(\gamma_{\bf G})M_{\bf k,p,q,[k+p-q]}^{(2)})\nonumber\\
 &\times& \left[\frac{B^{(4)}_{\bf k,p,q,[k+p-q]}}
   {(\epsilon_{\bf k}-\epsilon_{\bf p}-\epsilon_{\bf q}-\epsilon_{\bf k+p-q})}
   \mp\frac{{\rm sgn}(\gamma_{\bf G})B^{(6)}_{\bf k,p,q,[k+p-q]}}
   {\epsilon_{\bf k}+\epsilon_{\bf p}+\epsilon_{\bf q}+\epsilon_{\bf k+p-q}}
  \right].
 \label{eq.dyna2}
\end{eqnarray}
The upper (lower) signs correspond to the uniform (staggered) part.
The first-order correction Eq.~(\ref{eq.dyna1}) arises from the first term of
Eq.~(\ref{eq.f}).
In the second-order correction given by Eq.~(\ref{eq.dyna2}), 
the first term arises
from the first term of Eq.~(\ref{eq.f}), and the second term arises from 
the second term of Eq.~(\ref{eq.f}),
\[ G_{\alpha\alpha}^0({\bf k},\omega)\frac{1}{(2S)^2}
   \Sigma_{\alpha\beta}^{(2)}({\bf k},\omega)G_{\beta\beta}^0({\bf k},\omega)
  +G_{\beta\beta}^0({\bf k},\omega)\frac{1}{(2S)^2}
 \Sigma_{\beta\alpha}^{(2)}({\bf k},\omega)G_{\alpha\alpha}^0({\bf k},\omega).
\]
The third term of Eq.~(\ref{eq.dyna2}) is equivalent to
\[ \left.
  \frac{1}{(2S)^2}\frac{\partial\Sigma^{(2)}_{\alpha\alpha}({\bf k},\omega)}
 {\partial\omega}\right|_{\omega=\epsilon_{\bf k}}, \]
and arises from the second term of Eq.~(\ref{eq.f}),
\[ G_{\alpha\alpha}^0({\bf k},\omega)\frac{1}{(2S)^2}
  \Sigma_{\alpha\alpha}^{(2)}({\bf k},\omega)
  G_{\alpha\alpha}^0({\bf k},\omega).
\]
This is related to the second-order correction to the residue of 
the spin-wave pole in $G_{\alpha\alpha}({\bf k},\omega)$,
\[ \frac{1}{1- \left.
  \frac{1}{(2S)^2}\frac{\partial\Sigma^{(2)}_{\alpha\alpha}
 ({\bf k},\omega)}
 {\partial\omega}\right|_{\omega=\epsilon_{\bf k}}}
 \approx 
 1+ \left.
 \frac{1}{(2S)^2}\frac{\partial\Sigma^{(2)}_{\alpha\alpha}({\bf k},\omega)}
 {\partial\omega}
 \right|_{\omega=\epsilon_{\bf k}} .  \]
The fourth term arises from the third and fourth terms of Eq.~(\ref{eq.f}),
\[ I_{\mu\nu}({\bf k},\omega)G_{\nu\nu}^0({\bf k},\omega) 
 + G_{\mu\mu}^0({\bf k},\omega)\tilde I_{\mu\nu}({\bf k},\omega).
\]
No contribution arises from the last term of Eq.~(\ref{eq.f}),
$J_{\mu\nu}({\bf k},\omega)$.
Note that  the main momentum dependence around ${\bf k}=(0,0)$ arises from
the prefactors, $(\ell_{\bf k}+m_{\bf k})^2\propto \epsilon_{\bf k}$
and $(\ell_{\bf k}-m_{\bf k})^2\propto 1/\epsilon_{\bf k}$.

The three-spin-wave continuum arises only from the second-order corrections.
Note that the first term of Eq.~(\ref{eq.f}) has no contribution. 
After careful evaluation of other terms in Eq.~(\ref{eq.f}), we obtain
\begin{eqnarray}
 \rho_{u(s)}^{(2)}({\bf k},\omega) &=& 2S(\ell_{\bf k}\pm m_{\bf k})^2
  \frac{1}{(2S)^2}\left(\frac{2}{N}\right)^2\sum_{\bf pq}
  \delta(\omega-\epsilon_{\bf p}-\epsilon_{\bf q}-\epsilon_{\bf k+p-q})
  \nonumber\\
 &\times& \frac{1}{2}
 \ell_{\bf k}^2\ell_{\bf p}^2\ell_{\bf q}^2\ell_{\bf k+p-q}^2
 \Biggl[M_{\bf k,p,q,[k+p-q]}^{(1)}
     \pm{\rm sgn}(\gamma_{\bf G})M_{\bf k,p,q,[k+p-q]}^{(2)}\nonumber\\
 &-& \frac{2B^{(4)}_{\bf k,p,q,[k+p-q]}}
   {\epsilon_{\bf k}-\epsilon_{\bf p}-\epsilon_{\bf q}-\epsilon_{\bf k+p-q}}
   \mp\frac{2{\rm sgn}(\gamma_{\bf G})B^{(6)}_{\bf k,p,q,[k+p-q]}}
   {\epsilon_{\bf k}+\epsilon_{\bf p}+\epsilon_{\bf q}+\epsilon_{\bf k+p-q}}
  \Biggr]^2.
 \label{eq.conti}
\end{eqnarray}
The spectral shape may be modified by the renormalization of spin-wave
energies and by taking account of scattering spin waves due to mutual
interaction, which terms are present in $H_1$.
Therefore, it may be difficult to determine the spectral shape
in a consistent way with the $1/S$ expansion.
However, the total intensity, which is given by
\begin{equation}
 I_{u(s)}^{(2)}({\bf k})=\int_0^{\infty}{\rm d}\omega
             \rho_{u(s)}^{(2)}({\bf k},\omega),
\end{equation}
may be safely evaluated from Eq.~(\ref{eq.conti}).
Note that around ${\bf k}=(0,0)$ 
\begin{eqnarray}
 M_{\bf k,p,q,[k+p-q]}^{(1)}
     &\pm&{\rm sgn}(\gamma_{\bf G})M_{\bf k,p,q,[k+p-q]}^{(2)} \propto
     \epsilon_{\bf k}, \\
 -\frac{B^{(4)}_{\bf k,p,q,[k+p-q]}}
   {\epsilon_{\bf k}-\epsilon_{\bf p}-\epsilon_{\bf q}-\epsilon_{\bf k+p-q}}
   &\mp&\frac{{\rm sgn}(\gamma_{\bf G})B^{(6)}_{\bf k,p,q,[k+p-q]}}
   {\epsilon_{\bf k}+\epsilon_{\bf p}+\epsilon_{\bf q}+\epsilon_{\bf k+p-q}}
 \nonumber \\
 &\approx&
 \frac{B^{(4)}_{\bf k,p,q,[k+p-q]} 
   \mp{\rm sgn}(\gamma_{\bf G})B^{(6)}_{\bf k,p,q,[k+p-q]}}
   {\epsilon_{\bf p}+\epsilon_{\bf q}+\epsilon_{\bf k+p-q}} 
  \propto\epsilon_{\bf k},
\end{eqnarray}
and $(\ell_{\bf k}+m_{\bf k})^2\ell_{\bf k}^2\propto
const.$, $(\ell_{\bf k}-m_{\bf k})^2\ell_{\bf k}^2
\propto 1/\epsilon_{\bf k}^2$, we notice the dependences
around ${\bf k}=(0,0)$ as
$I^{(2)}_u({\bf k})\propto \epsilon_{\bf k}^2$,
$I^{(2)}_s({\bf k})\propto const.$ 

In the numerical evaluation of Eqs.~(\ref{eq.dyna2}) and (\ref{eq.conti}),
we sum up the values on $N_L^2$ points of ${\bf p}$ and ${\bf q}$ in the
first BZ, with $N_L=64$.

\subsection{Isotropic case}

\begin{figure}
\includegraphics[width=8.0cm]{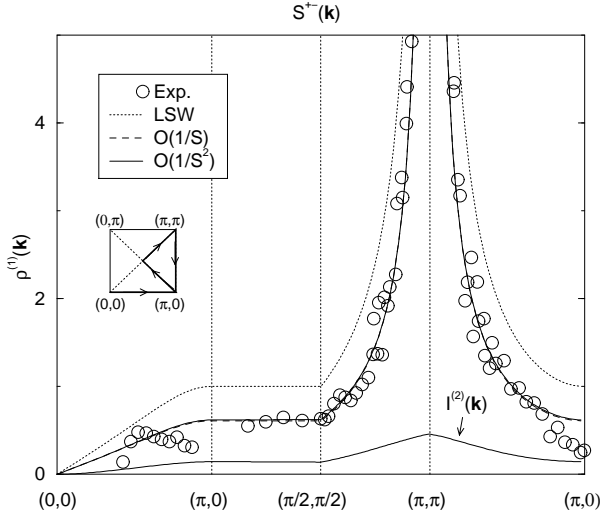}%
\caption{\label{fig.iso.struc}
Transverse dynamical structure factor as a function of momentum 
for the isotropic coupling. $S=1/2$.
The dotted, broken and solid lines represent the intensity of one spin-wave
excitation calculated within the LSW theory, the first-order correction, 
and the second-order correction, respectively. 
Thin solid line represents the intensity of three spin-wave excitations.
Experimental data are taken from Ref.~\onlinecite{Christensen04}.
Inset indicates high symmetry lines which momentum varies along.} 
\end{figure}

Figure \ref{fig.iso.struc} shows the spin-wave-peak intensity 
and the intensity of three-spin-wave continuum as a function of momentum 
along high-symmetry lines for $S=1/2$. 
Using the extended zone scheme, we assign the staggered part for line
$(0,0)-(\pi/2,\pi/2)$ to the values for line $(\pi,\pi)-(\pi/2,\pi/2)$ 
and also the staggered part for line $(0,0)-(\pi,0)$ to
the vales for line $(\pi,\pi)-(\pi,0)$.
The uniform part is assigned inside the first BZ.
At the zone boundary of the reduced BZ, the uniform and staggered parts
coincide with each other. 
The second-order corrections to the spin-wave-peak intensity becomes
one order of magnitude smaller than the first-order correction,
due to a cancellation among contributions of four terms 
in Eq.~(\ref{eq.dyna2}).
The intensity is almost determined within the first-order correction.
Thus the correction relative to the zero-th order value is independent 
of momentum.  We obtain around ${\bf k}=(0,0)$, 
\begin{equation}
 \rho^{(1)}_u({\bf k})=0.215 |{\bf k}|, \quad
 \rho^{(1)}_s({\bf k})= 1.72/|{\bf k}| \quad (S=1/2).
\end{equation}
These values should be compared with the $1/S$-expansion analysis
based on the Dyson-Maleev transformation,\cite{Canali93} 0.202 and 1.86.
A series expansion analysis by Singh\cite{Singh93} 
gives the values, 0.246 and 2.10, while a recent analysis by Zheng 
\textit{et al}.\cite{Zheng04} gives the values, 0.216 and 1.86.

The small second-order correction to $\rho^{(1)}_{u(s)}({\bf k})$
does not necessarily mean small three-spin-wave continuum.
At the zone boundary $(\pi,0)$, for example, we have 
$I^{(2)}(\pi,0)=0.143$ in addition to $\rho^{(1)}(\pi,0)=0.618$.
Such a considerable intensity of three spin-wave continuum has been
predicted in our previous paper\cite{Igarashi92-2} and 
others.\cite{Canali93,Zheng04}
It varies as proportional to $\epsilon_{\bf k}^2$ at the zone center,
and converges to a constant value at $(\pi,\pi)$.

\subsection{Anisotropic case}

\begin{figure}
\includegraphics[width=8.0cm]{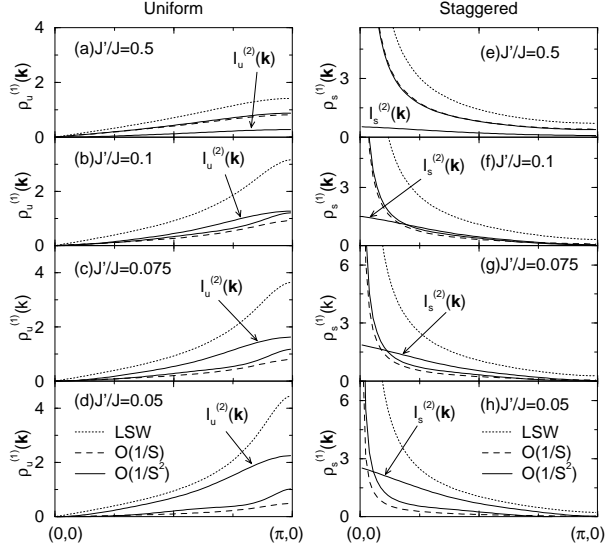}%
\caption{\label{fig.aniso.struc}
Transverse dynamical structure factor 
as a function of momentum along the symmetric line $(0,0)-(\pi,0)$
for anisotropic couplings;
(a)$J'/J=0.5$, (b)$J'/J=0.1$, (c)$J'/J=0.075$, and (d)$J'/J=0.05$. $S=1/2$.
The dotted, broken and solid lines represent the spin-wave-peak intensity
calculated within the LSW theory, up to the first-order correction, 
and up to the second-order correction, respectively. 
The thin solid line represents the three-spin-wave continuum intensity.}
\end{figure}

Figure \ref{fig.aniso.struc} shows the spin-wave-peak intensity 
and the intensity of three-spin-wave continuum along the symmetry line
$(0,0)-(\pi,0)$ for several anisotropic couplings. $S=1/2$.

The spin-wave-peak intensity is reduced by the first-order correction,
but is increased by the second-order correction.
Note that, although the residue of the spin-wave pole in the Green's function
is reduced by the self-energy, the other terms of the second-order
correction work to increase the intensity. The second-order correction 
increases with decreasing values of $J'/J$, but
the reduction due to the first-order correction is much larger than
the gain due to the second-order correction. As a result, the spin-wave-peak
intensity is strongly reduced. 

On the other hand, the intensity of three-spin-wave continuum increases
with decreasing values of $J'/J$.
It exceeds the spin-wave-peak intensity in the quasi-one dimensional situation.
Such large intensities of continuum spectra have been observed 
in the recent INS experiments on the quasi-one dimensional QHAF
such as KCuF$_3$\cite{Lake00} and BaCu$_2$Si$_2$O$_7$.\cite{Zheludev02}

\section{\label{sect.7} Concluding remarks}

We have systematically carried out the $1/S$ expansion up to the second order
on the basis of the HP transformation in the two-dimensional QHAF.
We have calculated the spin-wave energy in the whole BZ,
in comparison with the recent INS experiment 
for CFTD.\cite{Ronnow01,Christensen04}
We have found that the spin-wave energy at $(\pi,0)$ is about $2\%$
smaller than that at $(\pi/2,\pi/2)$ due to the second-order correction.
This is a correct tendency, but the value is somewhat smaller than
the experimental value $6\%$ and other theoretical estimates 
$7-9\%$.\cite{Singh95,Sandvik01,Zheng04}
We have so far not been able to find the reason for the difference, 
since the corrections higher than the second order of $1/S$ is expected
to be quite small. We have also calculated the transverse dynamical
structure factor. The second-order correction is found extremely small
in the one-spin-wave-peak intensity due to the cancellation
in the second-order terms, while it gives rise to substantial intensities 
of three-spin-wave continuum. This is consistent with previous 
studies.\cite{Igarashi92-2,Canali93,Singh95,Zheng04}

Canali and Wallin\cite{Canali93} reported that the value of the perpendicular 
susceptibility $\chi_{\perp}$ is different from the value 
in our previous paper\cite{Igarashi92-1} (for $J'/J=1$),
although they confirmed the same values for the second-order correction
to the sublattice magnetization $M_2$ and for the spin-wave velocity $V_x$.
The difference is not due to numerical errors, since the convergence 
with respect to $N_L$ has been carefully checked with changing 
$N_L=160, 320, 480$ (see Ref.~\onlinecite{Igarashi92-1}). 
As already discussed there, the value in Ref.~\onlinecite{Igarashi92-1}
satisfies the hydrodynamic relation, $V_x=(\rho_s/\chi)^{1/2}$ in an 
appropriate unit, with independently-evaluated spin-stiffness constant 
$\rho_s$.
As regards the dynamical structure factor, the difference
between the values by Canali and Wallin and the present values
might have the same origin as the difference in the perpendicular 
susceptibility, because the diagrams for the perpendicular susceptibility 
are closely related to those for the transverse dynamical structure factor.

The second-order correction is expected to become more important in 
quasi-one dimensional systems.
We have studied the crossover from two dimensions to one dimension
by weakening the exchange coupling in one direction.
All formulas are found formally the same as those for the isotropic coupling
with replacing $JSz$ by $2JS(1+\zeta)$.
It is found that the excitation energy is pushed up by the first-order and 
second-order corrections. With approaching the quasi-one dimensional situation,
the corrections make the excitation energy close to the des Cloizeaux-Pearson 
boundary in the one-dimensional QHAF for $S=1/2$. 
In the transverse dynamical structure factor,
the spin-wave-peak intensity is reduced by the first-order
correction, but is increased by the second-order correction.
The former exceeds the latter in the quasi-one dimensional situation, 
and thereby the peak intensity is strongly reduced from the LSW value.
On the other hand, the intensity of three-spin-wave continuum is found to 
become larger and exceeds the spin-wave-peak intensity.

In purely one-dimension, spin-one excitations are considered excitations
of two spinons of a spin-one-half excitation.
In this respect, the spin wave might be considered 
as a bound state of two spinons.  Our finding that the weight 
$\rho_1({\bf k})$ of the spin-wave peak decreases with $J'/J\to 0$
is consistent with this picture. Large intensities
of three-spin-wave continuum might be replaced by two-spinon continuum
in the one-dimensional limit. 
Recently, INS experiments have been carried out at low temperatures
in the quasi-one dimensional systems such as KCuF$_3$\cite{Lake00} 
and BaCu$_2$Si$_2$O$_7$,\cite{Zheludev02} and large broad spectra 
have been observed in addition to a peak in the transverse dynamical structure 
factors. This behavior as well as the behavior of the longitudinal component
have been analyzed by the chain-mean-field and random phase 
approximation.\cite{Essler97,Zheludev03}
It may be interesting to analyze these data in terms of the $1/S$ expansion
by starting from a detailed three-dimensional model 
with directional anisotropy.

\begin{acknowledgments}
J.I. thanks MPI-PKS at Dresden for hospitality during his stay, 
where this work started.
This work was partially supported by a Grant-in-Aid for Scientific Research 
from the Ministry of Education, Science, Sports and Culture, Japan.

\end{acknowledgments}


\bibliography{paper2}

\end{document}